\documentclass{article}

\usepackage{arxiv}

\usepackage[utf8]{inputenc} 
\usepackage[T1]{fontenc}    
\usepackage{hyperref}       
\usepackage{url}            
\usepackage{booktabs}       
\usepackage{amsfonts}       
\usepackage{nicefrac}       
\usepackage{microtype}      
\usepackage{lipsum}
\usepackage{amsmath}
\usepackage{inputenc}

\usepackage{subcaption}
\usepackage{rotating}
\usepackage{tikz}
\usepackage{lscape}
\usetikzlibrary{positioning}
\usepackage{array,calc}
\newlength\lena \newlength\lenb \newlength\lenc \newlength\lend
\newcolumntype{P}[1]{>{\centering\arraybackslash}p{#1}} 

\title{Greenhouse Gas Emission Prediction on Road Network using Deep Sequence Learning}

\author{
	\noindent
  Lama Alfaseeh \\
  Laboratory of Innovations in Transportation (LiTrans)\\
  Ryerson University\\
  Toronto, Canada \\
  \texttt{lalfaseeh@ryerson.ca}
   \And
  Ran Tu\\
  University of Toronto\\
  Toronto, Canada\\
 \texttt{ran.tu@mail.utoronto.ca}
\And   
  Bilal Farooq \\
  Laboratory of Innovations in Transportation (LiTrans)\\
  Ryerson University\\
  Toronto, Canada \\
  \texttt{bilal.farooq@ryerson.ca}
  \And
  Marianne Hatzopoulou\\
  University of Toronto\\
  Toronto, Canada\\  
  \texttt{marianne.hatzopoulou@utoronto.ca} \\
  \\
}

\begin{document}
\maketitle

\begin{abstract}
Mitigating the substantial undesirable impact of transportation systems on the environment is paramount. Thus, predicting Greenhouse Gas (GHG) emissions is one of the profound topics, especially with the emergence of intelligent transportation systems (ITS). We develop a deep learning framework to predict link-level GHG emission rate (ER) (in $\text{CO}_{2\text{eq}}$ gram/second) based on the most representative predictors, such as speed, density, and the GHG ER of previous time steps. In particular, various specifications of  the long-short term memory (LSTM) networks with exogenous variables are examined and compared with clustering and the autoregressive integrated moving average (ARIMA) model with exogenous variables. The downtown Toronto road network is used as the case study and highly detailed data are synthesized using a calibrated traffic microsimulation and MOVES. It is found that LSTM specification with speed, density, GHG ER, and in-links speed from three previous minutes performs the best while adopting 2 hidden layers and when the hyper-parameters are systematically tuned. Adopting a 30 second updating interval improves slightly the correlation between true and predicted GHG ERs, but contributes negatively to the prediction accuracy as reflected on the increased root mean square error (RMSE) value. Efficiently predicting GHG emissions at a higher frequency with lower data requirements will pave the way to non-myopic eco-routing on large-scale road networks {to alleviate the adverse impact on the global warming}. 
\end{abstract}




\section{Introduction}
\label{Introduction}
Transportation systems have been consistently ranked as the first largest source of GHG emissions {(mainly carbon dioxide)} in the U.S. producing 29\% of the the total emissions \cite{epa2017sources}. {GHG emission is among the main contributors to the global warming and climate change \cite{liu2020predictions}.} Hence, topics related to GHG reduction have grabbed the attention of transportation researchers in the past few decades. On the solution side, the employment of the intelligent transportation systems (ITS) has been considered as a most favourable approach \cite{zegeye2009model} to alleviate the undesirable impact of transportation systems on the environment. The ability of the ITS to capture high resolution information about traffic conditions contributes to a more efficient traffic management process. Furthermore, the availability of real-time data allows the shift to non-myopic routing for {more sustainable transportation systems}.

Although prediction has been considered as a crucial topic, most of the studies tackled the matter from an aggregated level. In other words, GHG emissions are modelled or predicted at the national level. Examples include \cite{pao2011modeling}, \cite{lin2011grey}, \cite{antanasijevic2014forecasting}, and \cite{radojevic2013forecasting}. The main justification is the scarcity of microscopic data and complexity associated with the models. 
With regards to the models used, Grey Models (GMs) \cite{dengiz2018grey}, the Auto Regressive Integrated Moving Average (ARIMA) \cite{rahman2017modeling}, and the Artificial Neural Networks (ANNs) \cite{abdullah2015methods} were widely used to forecast GHG emissions. GMs were commonly used due to the fact that they require comparatively small number of data points and can manage the case of limited or missing data \cite{dengiz2018grey}, which can affect adversely the accuracy. ARIMA models are based on historical values of the predicted variable, such as CO$_2$ \cite{rahman2017modeling}. It is worth mentioning that ARIMA models are associated with a major limitation related to their assumption of linear relationship between variables \cite{zhang2003time}, which restricts ARIMA application in the case of complex non-linear time series. Thus, ANNs have been introduced to incorporate the non-linear relationship between predictors and responses. It has been shown that NNs outperform other models when they are adopted to predict the environmental pollutants \cite{singh2012linear}. Recurrent neural networks (RNNs), which are a type of the deep neural networks (DNNs), were introduced and were associated with multiple layers between the input and output layers, unlike the artificial neural networks (ANNs). The RNNs define the correct mathematical manipulation to give output from an input, whether it is a linear relationship or a non-linear relationship \cite{aggarwal2018neural}. The Long-Short Term Memory (LSTM) network is a category of the RNNs, that overcomes the drawback of regular RNNs, the vanishing gradient problem \cite{amarpuri2019prediction}. The LSTM network has been considered as one of the most powerful RNN architectures where sequential data is involved \cite{lipton2015critical}. The LSTM deep network consists of three main components: 1) input gate, which controls the information fed to the network, 2) forget gate, which controls whether to keep or forget the information of the previous time step, and 3) output gate, which controls what information to give out of the network \cite{amarpuri2019prediction}. With reference to the LSTM application, authors mainly utilized it to predict at an aggregated level, spatially and temporally, as in \cite{ameyaw2019investigating}, \cite{ameyaw2018analyzing}, and \cite{huang2019grey}. 

While the existing literature forecast GHG emissions at an aggregated level, utilizing generally fuel and economical factors, this study aims to predict GHG ER (in $\text{CO}_{2\text{eq}}$ g/sec) at a link level using a deep learning framework, based on LSTM with exogenous variables, while using microscopic data points. An agent-based traffic simulation is utilized to obtain traffic and environmental high resolution information. Furthermore, a comparison is conducted with two other models: ARIMA with exogenous variables, and clustering. The impact of utilizing a deeper LSTM model while systematically tuned is examined. To evaluate the performance, we use the correlation coefficient between observed and predicted GHG ERs (in $\text{CO}_{2\text{eq}}$ g/sec), the fit to the ideal straight curve reflecting on the precision, R$^2$ statistics, and the root mean square error (RMSE) reflecting on the accuracy. Finally, a crucial aspect is investigated related to defining the impact of the updating interval of the predictors when LSTM is adopted. Predictors at every 30 seconds and 1 minute are used for the comparison and trade offs are illustrated.
\par The main contributions of this work are as follows:
\begin{enumerate}
	\item Development of a deep learning framework based on LSTM to predict GHG ER (in $\text{CO}_{2\text{eq}}$ g/sec) at link level in a highly congested urban network, while utilizing microscopic data.
	\item Systematic analysis of the importance of various predictors contributing to the models.
	\item Comparison of the results with the two commonly used emission prediction models, ARIMA and clustering, and demonstration of the strengths and limitations associated with each.
	\item Illustration of the impact of utilizing different lengths of time interval for predictors and demonstrating the effect of the systematic tuning while applying LSTM.
\end{enumerate}

This work is organized as follows: Section \ref{Literature Review} presents a brief literature review of studies that predicted GHG. The methodology related to the predictive models utilized is in Section \ref{Methodology}. The description of the case study is illustrated in section \ref{Case study}. The traffic simulation and the emission model for collecting data are presented in section \ref{Data collection} including the details of the scenarios simulated. Discussion and results are in Section \ref{Discussion and results}. Finally, concluding remarks and future outlook are in section \ref{Conclusion}.

\section{Literature Review}
\label{Literature Review}
A large number of studies predicted GHG or CO$_2$ at national level using yearly data points of predictors while adopting different models. \cite{pao2011modeling} predicted CO$_2$ emissions based on income and energy consumption by using grey prediction model (GM) in Brazil. The authors compared between GM and ARIMA model and found that results are comparable in terms of the forecasting performance. Another study by \cite{lin2011grey} used GM to predict CO$_2$ emissions in Taiwan. ARIMA model was used by \cite{rahman2017modeling} and \cite{tudor2016predicting} to predict CO$_2$ emissions in Bangladesh and Bahrain, respectively. \cite{antanasijevic2014forecasting} predicted GHG emissions at national level, of European countries, employing a new approach based on ANNs. The predictors considered were agriculture, transportation, energy supply and use, and waste. Another study by  \cite{radojevic2013forecasting} utilized ANN to predict CO$_2$ emissions in Serbia based on the share of renewable sources of energy, the gross domestic product, the gross energy consumption, and energy intensity were selected as the input parameters \cite{radojevic2013forecasting}. \cite{sun2017factor} predicted CO$_2$ emissions by utilizing extreme learning machine (ELM), which is a type of ANNs, based on particle swarm (PSO) optimization. The authors used the variables impacting the produced CO$_2$ and classified them into one group. In addition, they found that their proposed approach outperformed the ELM and the back propagation neural network in terms of the RMSE and MAPE. \cite{grote2018practical} developed Practical Emissions Model for Local Authorities emission (PEMLA) to be used by local government authorities. PEMLA estimates CO$_2$ at a network level based on data collected from inductive loop detectors, which are installed as part of urban traffic control systems. Five traffic variables were taken into consideration: 1) traffic average speed (km/h); 2) traffic density (vehicles/km); 3) traffic average delay rate (seconds/vehicle·km); 4) access density (intersections/km); and 5) the square of traffic average speed (km/h)$^2$. The authors employed Multiple Linear Regression (MLR) and the ANN to define the relationship between variables and predict CO$_2$, respectively. The case study was Southampton, a city on the South coast of the UK with a population of approximately 255,000 and 5 min traffic variables were obtained from detectors to estimate network level CO$_2$ emissions \cite{grote2018practical}.

The Long-Short Term Memory (LSTM) network has been considered as one of the best RNN architectures for the case of sequential data \cite{lipton2015critical}. With regards to the LSTM application, \cite{ameyaw2019investigating} utilized it to predict country level CO$_2$. The authors considered several predictors based on their correlation with CO$_2$. Gross fixed capital formation which was measured as a percentage of gross domestic product, total labor force, and trade were also adopted as predictors in addition to gross domestic product per capita \cite{ameyaw2019investigating}. Similarly, \cite{ameyaw2018analyzing} predicted CO$_2$ in five West African countries, but considering only one predictor which is the gross domestic product. \cite{huang2019grey} predicted CO$_2$ in China based on four predictors using LSTM. Clustering is another technique used for prediction in transportation systems, examples include \cite{poucin2018activity,Ran2018TRB} and \cite{gmira2017travel}. In some cases, clustering is used in combination with other models, such as MapReduce Framework as in \cite{zhao2011mapreduce}. 

Unlike the aforementioned studies above, which can be classified as top-down approaches, \cite{dong2019carbon} developed a predictive model for diesel trucks and gasoline passenger cars based on volume to capacity (V/C) ratio as an explanatory variable. Due to time and cost restrictions, V/C range was taken only between 0.15 and 1.1. The authors applied the predictive model on an actual expressway in China \cite{dong2019carbon}. The authors defined a regression curve that illustrated the relationship between emissions and V/C \cite{dong2019carbon}. The main shortcoming of their approach is that the density and speed of congested traffic conditions are not captured efficiently when V/C is the only predictive variable and when values out of the defined range are not considered. Hence, their approach did not represent the real situation of traffic and considered a small case study. \cite{wang2015fine} developed a model to predict 5-min series of carbon monoxide (CO) and fine particulate matter (PM2.5) of an intersection in Shanghai. The proposed model is a hybrid model combining wavelet neural network and genetic algorithm (GA-WNN). In terms of the traffic predictors, delay and queue length of different directions were taken into consideration in addition to other predictors \cite{wang2015fine}. Another study by \cite{zhao2011mapreduce} suggested a MapReduce framework for on-road mobile fossil fuel combustion CO$_2$ emissions estimation. Data was obtained from ITS and processed to estimate GHG emissions. Their model had three layers; infrastructure, distributed computing platform, and application layer. In the application layer, data was clustered, outliers were removed, and GHG emissions were estimated based on ERs obtained from speed on links. It is worth mentioning that this approach estimated emission based on aggregated data using regression models of speed \cite{zhao2011mapreduce}. One more study by \cite{coelho2009numerical} developed a traffic and emission decision support (TEDS) tool to provide an overall pollution estimate for a traffic interruption. The authors defined the relationship between several pollutants, carbon monoxide (CO), carbon dioxide (CO2), nitric oxide (NO), and hydrocarbons (HC), and queue length and stops. It was assumed that average speed and flow were constant of zones studied, which is a limitation as realistic conditions were not captured. Their case study was two highways to the city of Lisbon (Portugal), namely Freeway A5 and Highway N6 \cite{coelho2009numerical}.
\par To sum it up, in the existing literature, GHG emissions are mostly predicted at an aggregated level, spatially and temporally, and based mainly on yearly data points of fuel consumption, gross domestic product, or other economical factors. Although transportation systems contribute substantially to the GHG produced \cite{luo2016real}, there is a lack of studies tackling the GHG prediction matter at a microscopic level that will be used in several controlling and management processes, such as routing for more sustainable transport systems. To the best of our knowledge, this study is the first of its kind that employs a deep learning approach i.e. LSTM, to predict GHG ER (in $\text{CO}_{2\text{eq}}$ g/sec) at link level for every minute or less, using the most representative predictors for a congested urban network.

\section{Methodology}
\label{Methodology}
Before discussing the LSTM methodology in detail, we describe the two dominantly used models, ARIMA and clustering, for comparison purposes. ARIMA is a statistical model, while clustering is a machine learning algorithm.

\begin{figure}[!ht]
	\centering
	\begin{subfigure}[t]{.45\textwidth}
		\includegraphics[width=3in]{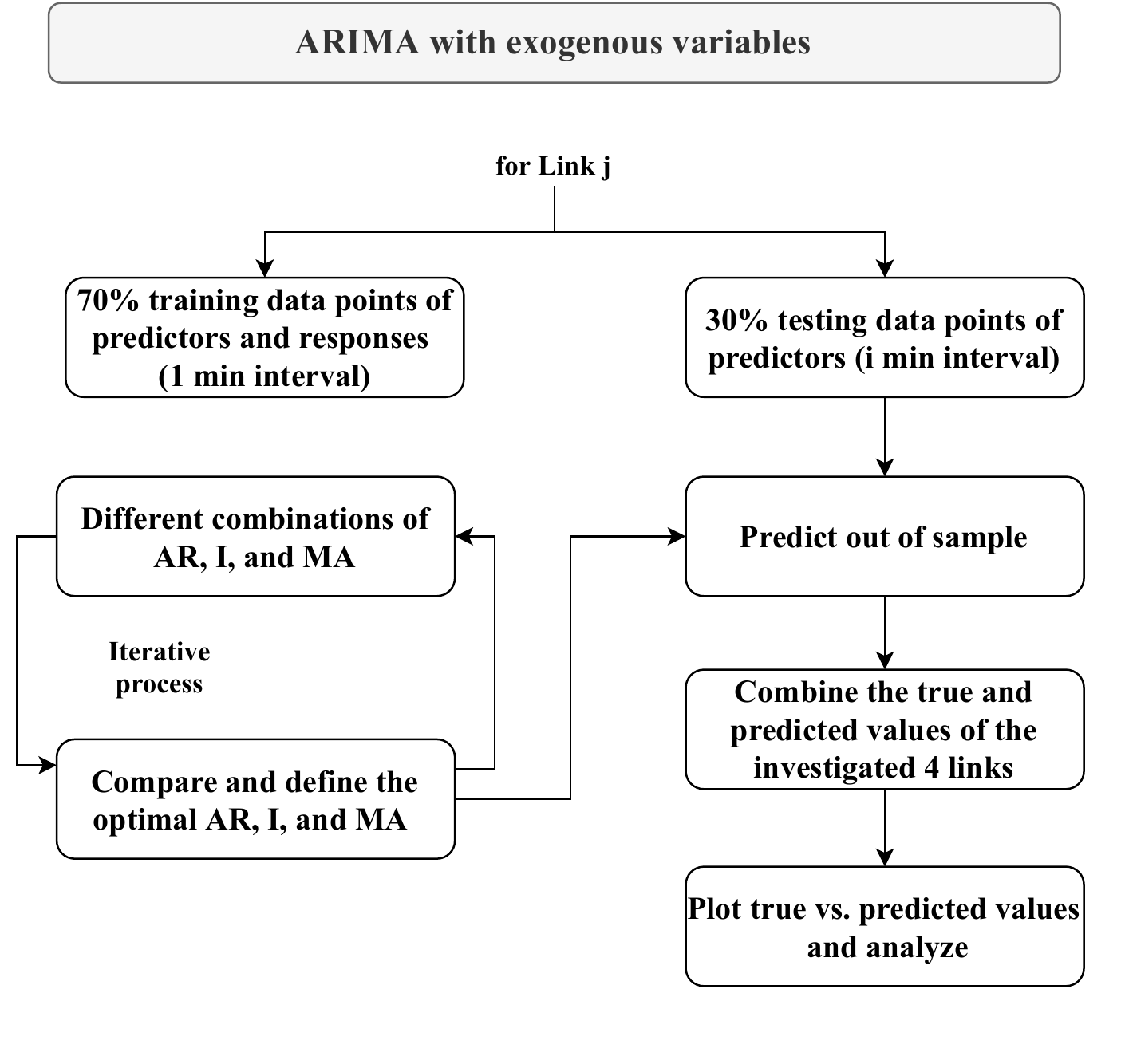}
		\label{ARIMA_methodology}
	\end{subfigure}
	\begin{subfigure}[t]{.45\textwidth}
	\includegraphics[width=3in]{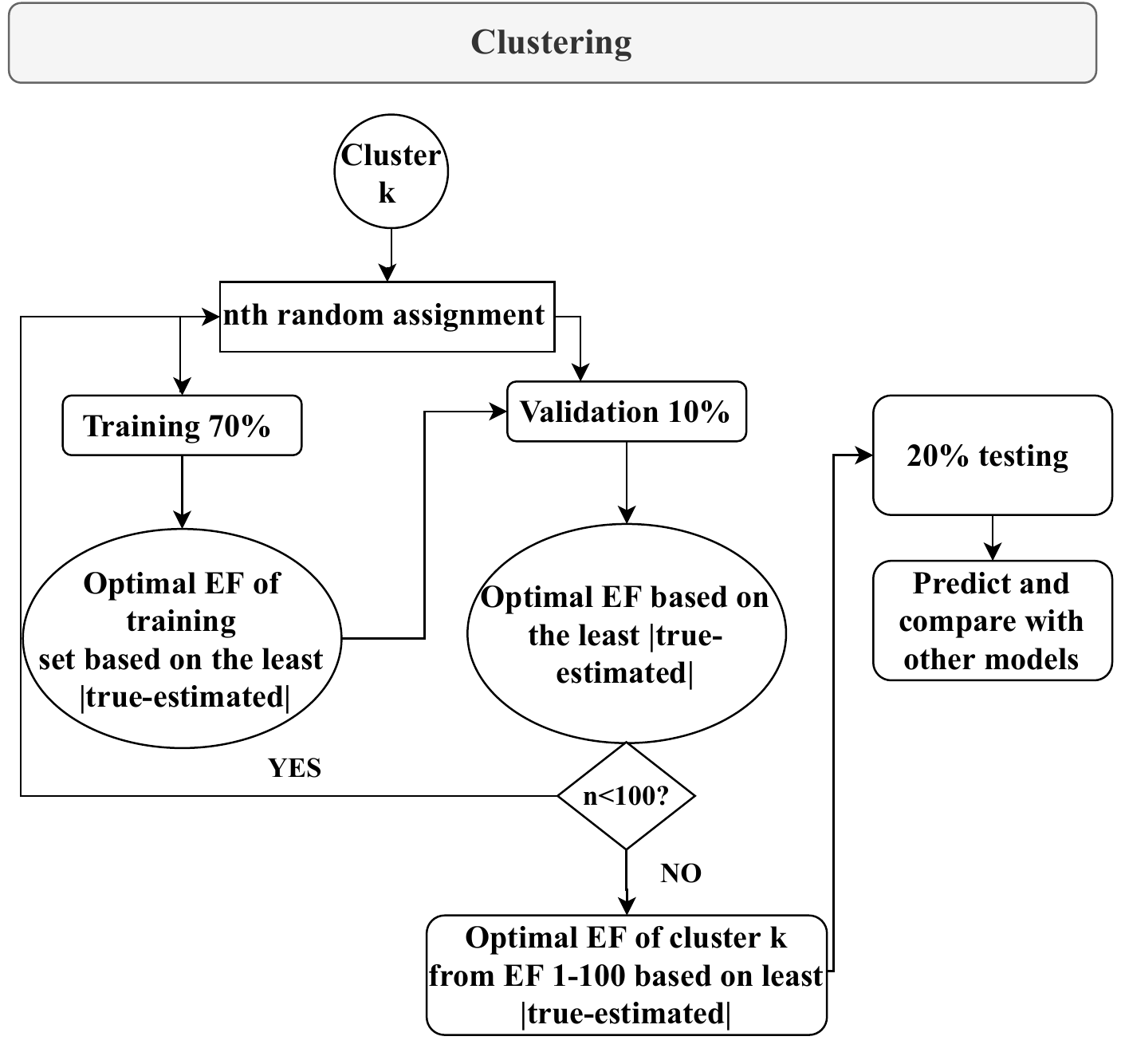}
		\label{Clustering_methodology}
		\end{subfigure}\\
	\hfil
	\centering
	\begin{subfigure}[t]{\textwidth}
		\centering
	{\includegraphics[width=6in]{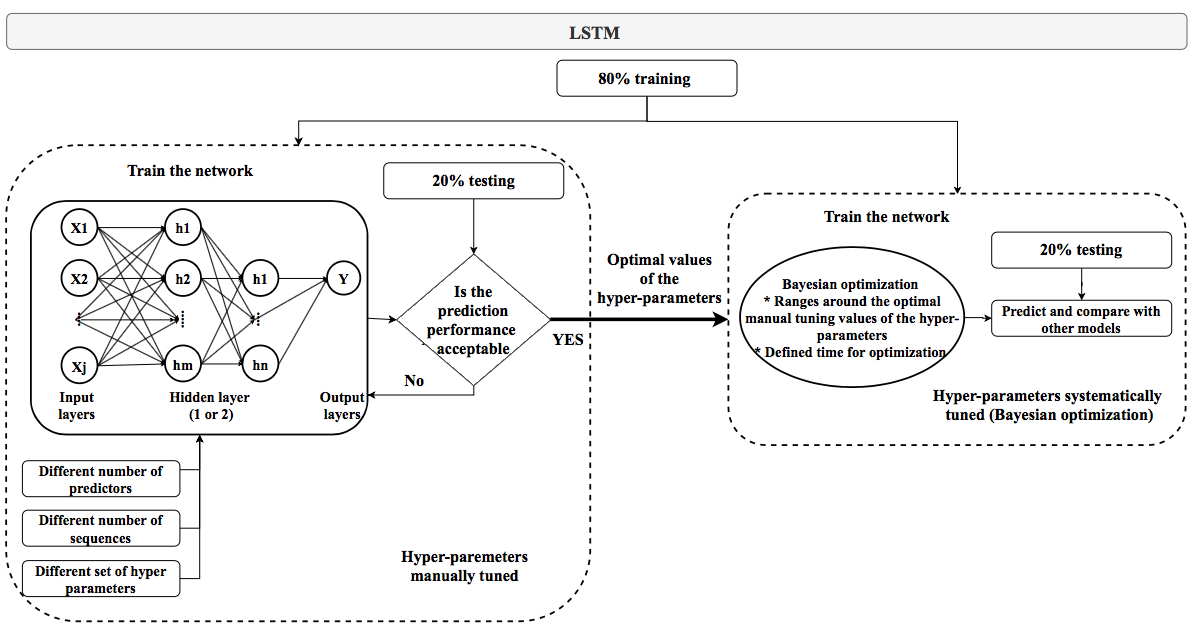}}
		\label{LSTM_final_2}
	\hfil
		\end{subfigure}
	\caption{Methodology for a) ARIMA with exogenous variables, b) clustering, and c) LSTM with exogenous variables}
	\label{methods}
\end{figure}

\subsection{ARIMA with exogenous variables}
\label{ARIMAX_methodology_text}
ARIMA (p,d,q) has traditionally been used for time series data that can be made to be ``stationary'' by differencing of ``d'' order. In other words there should be no trends in the data in order to predict using the ARIMA model efficiently. Where: “p” and “q” are small integers reflecting on the “autoregressive” (AR) and “moving-average” (MA), respectively. To include $r$ exogenous variables while using ARIMA, Equation \ref{equation_ARIMAX} is applied.
\begin{equation}
	\hat{y_t}= \mu + \sum_{i=1}^{p}\phi_i y_{t-i} + \sum_{k=1}^{r}\beta_k x_{tk} + e_t+ \sum_{j=1}^{q}\theta_j e_{t-j}
	\label{equation_ARIMAX}
\end{equation}
Where $\hat{y_t}$ is the response at time $t$, $\mu$ is the constant, $\phi_i$ is the AR coefficient at lag $i$, $y_{t-i}$ is the value of the variable in concern for prediction at time $t-i$, $\beta_k$ is a coefficient of exogenous variable $k$, $x_{tk}$ is exogenous variable $k$ at time $t$, $\theta_j$ is the MA coefficient at lag $j$, and $e_{t-j}=y_{t-j}-\hat{y}_{t-j}$ is the forecast error that was made at period $t-j$ \cite{box2015time}.


Figure \ref{methods} illustrates the steps followed to develop the ARIMA model and predict the GHG ERs. A ratio of 70\% to 30\%, training to testing, is considered. To define the optimal parameters (p,d,q), an iterative process took place while considering the auto-correlation, partial auto-correlation plots of the differenced series, and root unit which is a measure that signalizes when the time series is under or overdifferenced. One of the major shortcomings of ARIMA models in our context is that a separate model needs to be estimated for each link, based on the link data, to assure that the time series is stationary. In other words, ARIMA lacks the spatial dimension.


\subsection{Clustering}
\label{Clustering_methdology_text}
Clustering is an important tool in data mining applications. It mainly groups objects so that in one group the objects are more related to each other than those in other groups \cite{mann2013review}. To classify the traffic conditions at link level using the most important variables, K-mean clustering has been utilized. It aims to classify data points into K clusters. Every data point belongs to a cluster with the  minimum distance to the centroids of that cluster. In K-mean clustering, the optimal cluster is defined when the total intra-cluster variance, or, the squared error is minimized \cite{poucin2018activity} following Equation \ref{equation_clustering}.
\begin{equation}
	M= \sum_{m=1}^{k} \sum_{i=1}^{n} || x_i^{(m)}-c_m ||^2
	\label{equation_clustering}
\end{equation}
Where $M$ is the objective function, $k$ is the number of clusters, $n$ is the number of data points (observations), $x_i^{(m)}$ is the observation $i$ being tested for cluster $m$, $c_m$ is the centroid of cluster $m$.
The number of clusters should be defined based on the data and a statistical analysis. To define the optimal number of clusters, the elbow method/sum of squared error \cite{poucin2018activity}, which measures the sum of squared distances between the points within a cluster is the guide. {Figure \ref{methods}} presents the steps followed. For clustering, dataset has been divided into 70\%, 10\%, and 20\% for training, validating, and testing, respectively \cite{tu2018development}. The optimal GHG ER (g/sec) of each cluster is defined based on the minimum sum of absolute distances between GHG ERs and the centroids of the examined cluster.

\subsection{LSTM with exogenous variables}
\label{LSTM_methodology_text}
The main focus of this work is to develop an LSTM based learning framework and compare it to already established models including ARIMA and clustering. LSTM is associated with three gates \cite{hochreiter1997long}, input, forget, and output following the below Equations \ref{equation_LSTM1}, \ref{equation_LSTM2}, and \ref{equation_LSTM3}, respectively.
\begin{equation}
	i_t=\sigma (w_i[h_{t-1},x_t] + b_i)
	\label{equation_LSTM1}
\end{equation}

\begin{equation}
	f_t=\sigma (w_f[h_{t-1},x_t] + b_f)
	\label{equation_LSTM2}
\end{equation}

\begin{equation}
	o_t=\sigma (w_o[h_{t-1},x_t] + b_o)
	\label{equation_LSTM3}
\end{equation}
Where: $i_t$ represents the input gate, $f_t$ represents the forget gate, $o_t$ represents the output gate, $\sigma$ represents the sigmoid function, $w_x$ represents the weight for gate $x$ neurons, $h_{t-1}$ represents the output of the previous LSTM block at time $t-1$, $x_t$ represents the input at current time step $t$, and $b_x$ represents biases for respective gates ($x$).

Defining the best network is an iterative process and is dependent on the selection of the predictors, number of sequences, and set of hyper-parameters \cite{reimers2017optimal, hutter2015beyond}. It is worth mentioning that increasing the depth of NNs \cite{hermans2013training, pascanu2013construct} and effectively tuning the network \cite{snoek2012practical} may contribute positively to the prediction performance. Thus, one and two hidden layers are investigated in this work and Bayesian optimization \cite{wu2019hyperparameter} is utilized for the systematic tuning. With reference to hyper-parameters tuning procedure, two stages have taken place. The first is the manual, while, the second is the systematic. {Figure \ref{methods}} presents the methodology followed in LSTM application. A ratio of 80\%-20\% has been considered for training to testing, respectively.  To compare the performance of LSTM to ARIMA and clustering, four indicators are employed, the correlation coefficient between observed and predicted GHG ERs (in $\text{CO}_{2\text{eq}}$ g/sec), the fit to the ideal straight curve, the R$^2$, and the RMSE.

\section{Case study}
\label{Case study}
\begin{figure*}[ht]
	\centering
	\includegraphics[width=\textwidth]{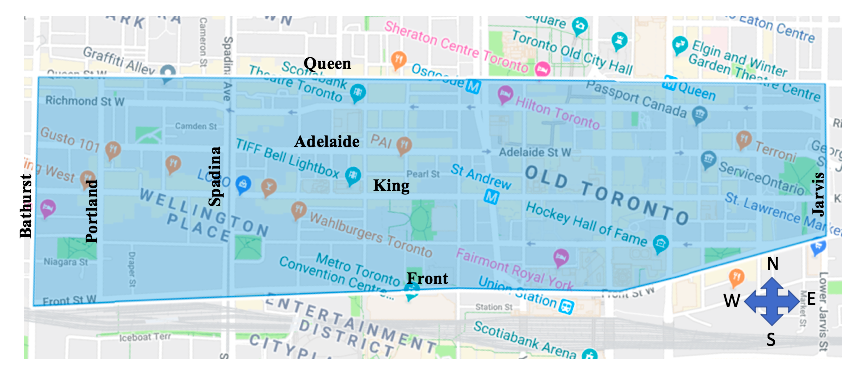}
	\caption{Case study, downtown Toronto}
	\label{Case_study}
\end{figure*}

The road network of downtown Toronto is used as the case study. It was selected due to its high level of recurrent congestion during the morning peak period. The network consists of 223 links and 76 nodes. Links have different characteristics in terms of the number of lanes, free flow speed, and number of directions to assure heterogeneity. {Figure \ref{Case_study}} illustrates the area, including the major roads. 

\begin{figure}[!h]
	\centering
	
	\begin{subfigure}[t]{.45\textwidth}
		{\includegraphics[width=2.5in]{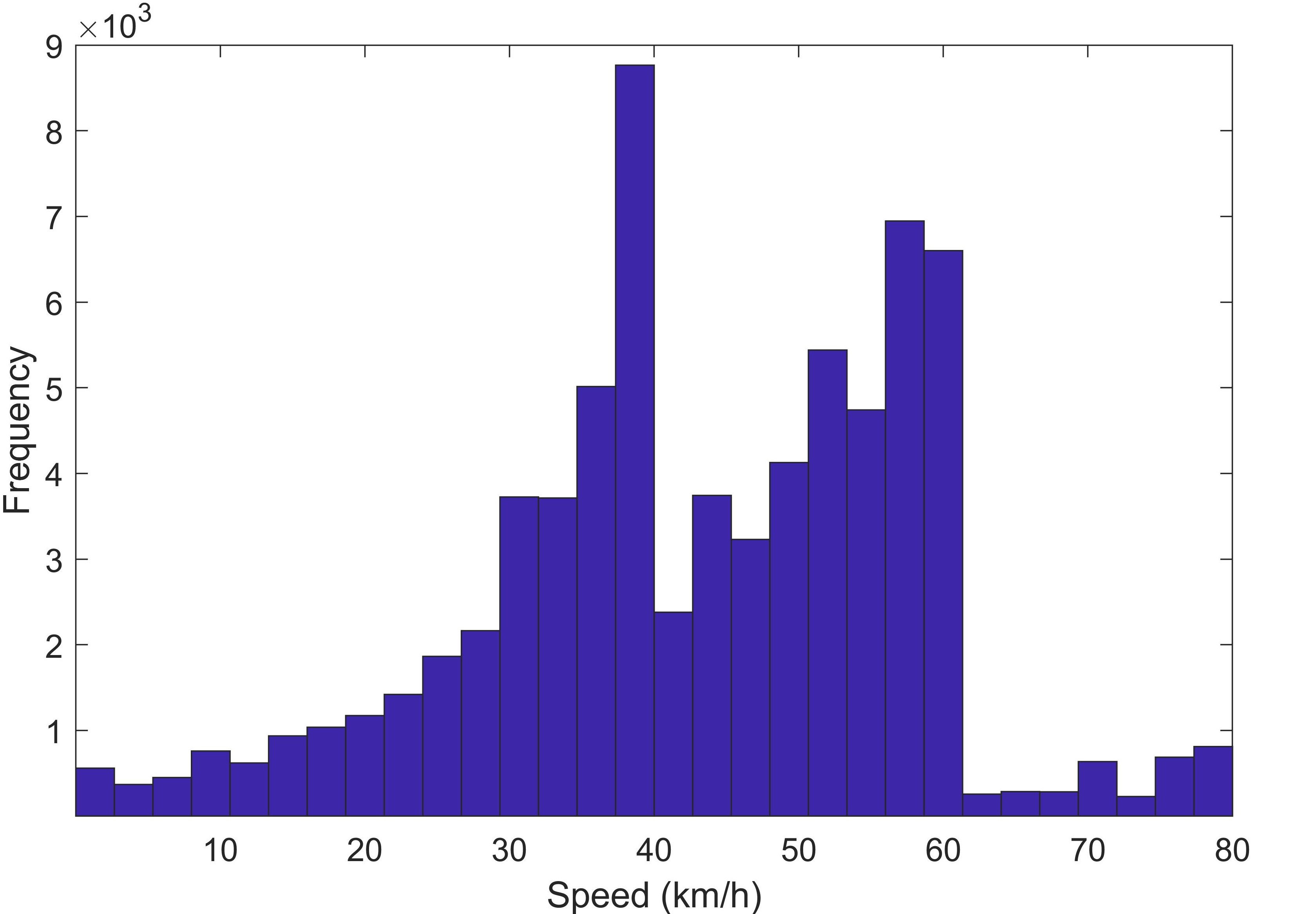}
		\label{histogram speed}}
	\end{subfigure}
	\hfil
	\begin{subfigure}[t]{.45\textwidth}
		{\includegraphics[width=2.5in]{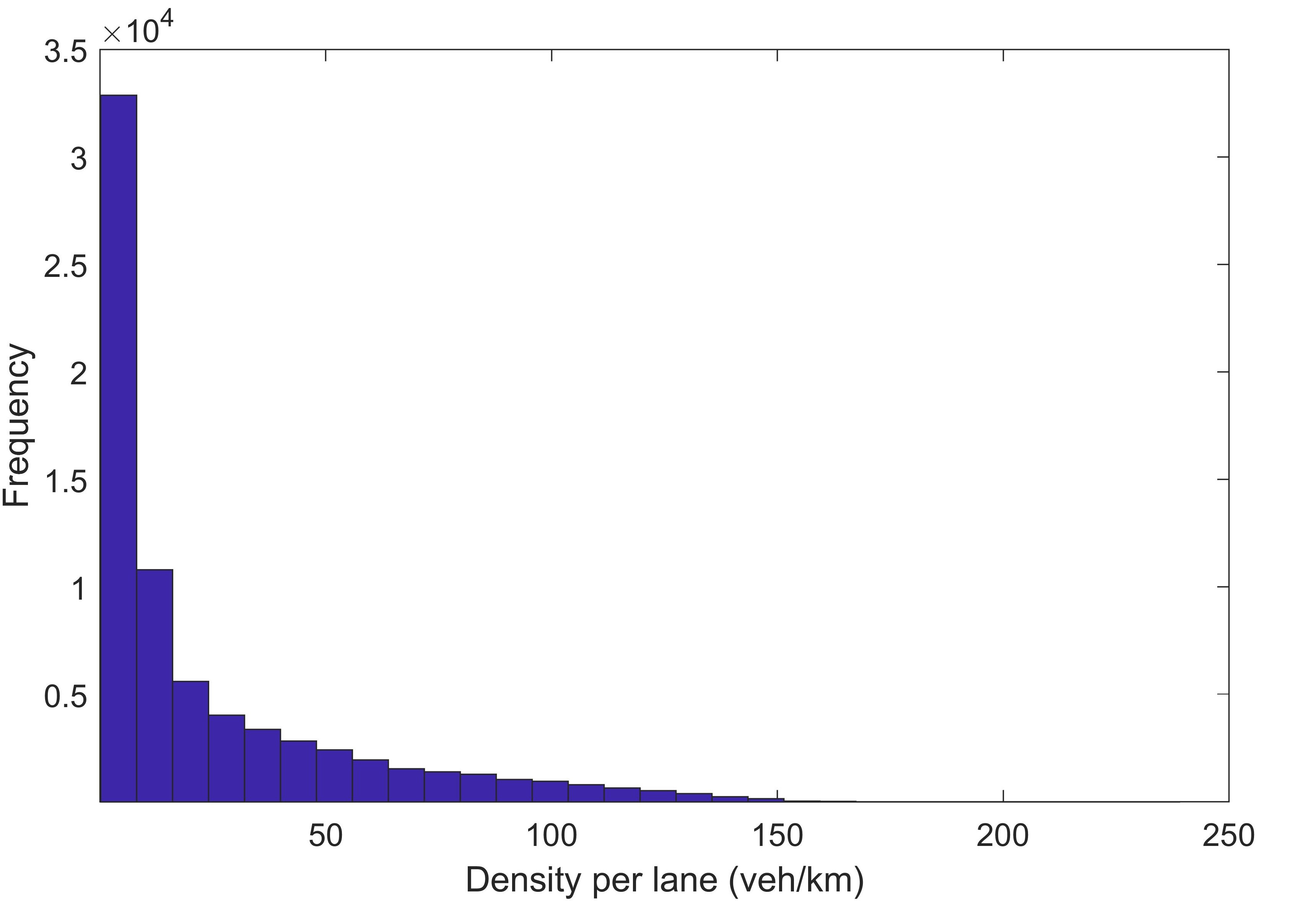}
		\label{histogram density per lane}}
	\end{subfigure}
	\hfil
	\begin{subfigure}[t]{.45\textwidth}
		{\includegraphics[width=2.5in]{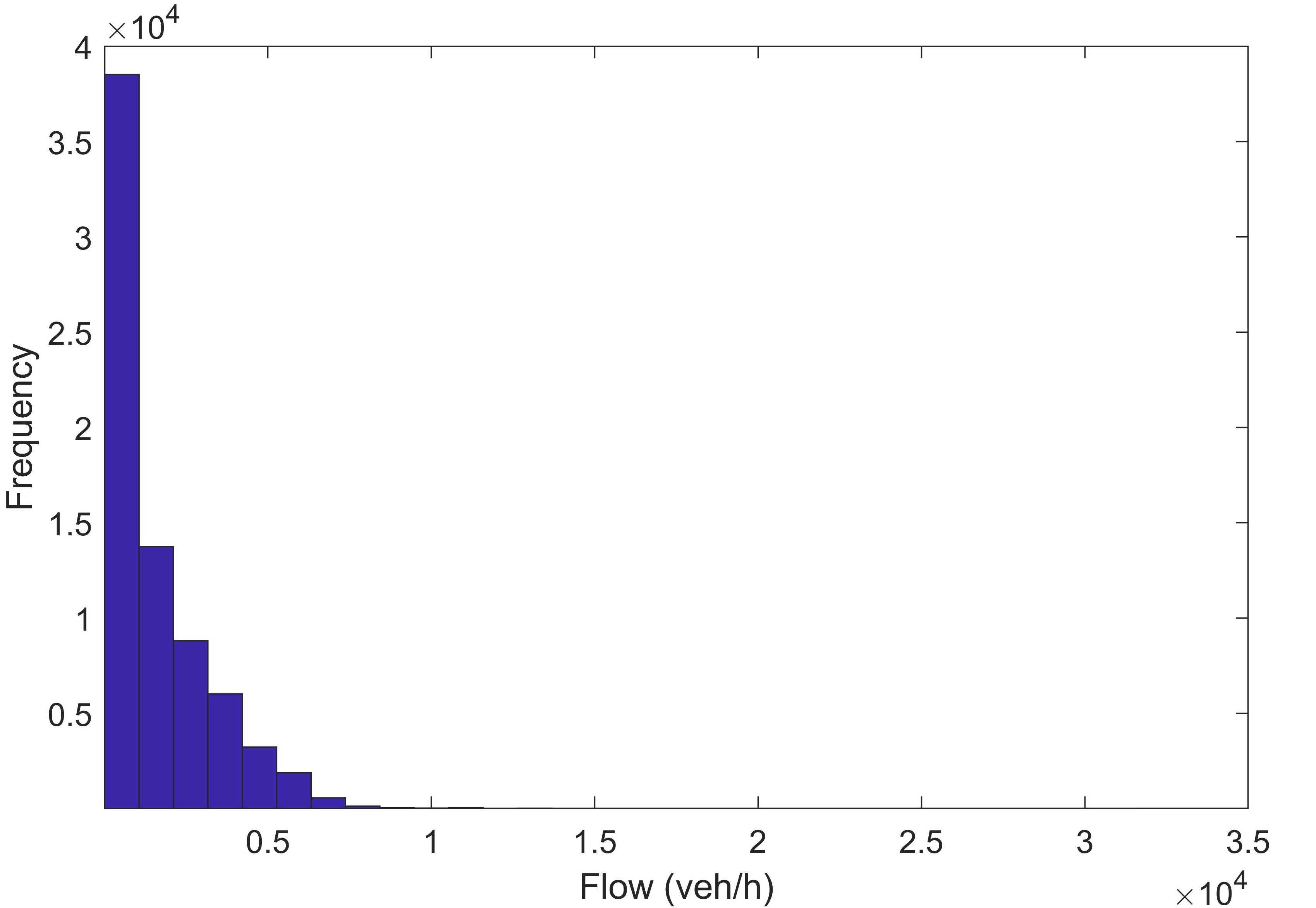}
		\label{histogram total flow}}
	\end{subfigure}
	\hfil
	\begin{subfigure}[t]{.45\textwidth}
		{\includegraphics[width=2.5in]{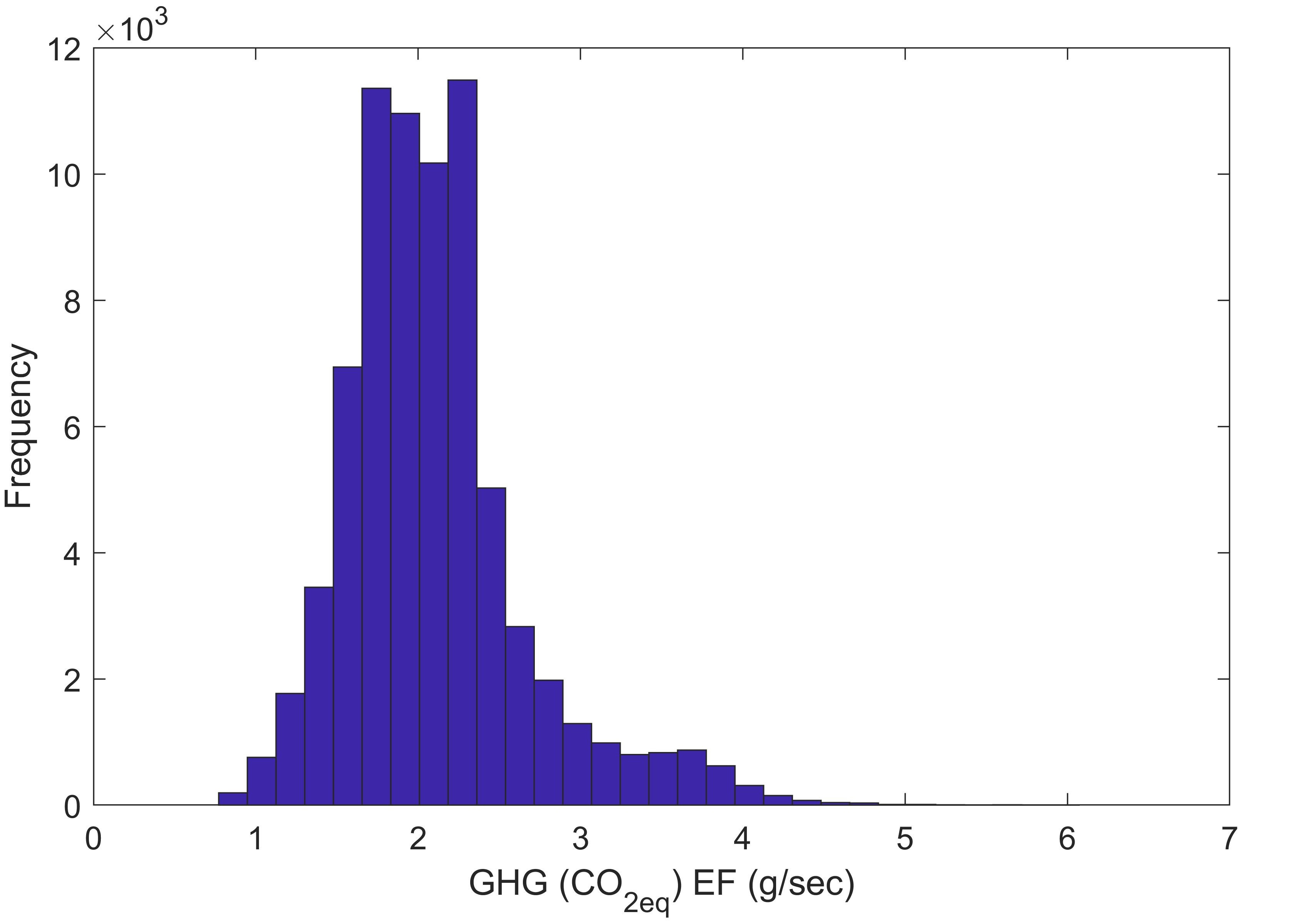}%
		\label{histogram CO2}}
	\end{subfigure}
	\hfil
	\caption{Histogram of a) speed, b) density per lane, c) flow, and d) GHG ERs (in $\text{CO}_{2\text{eq}}$ g/sec) }
	\label{histograms}
\end{figure}

\section{Data collection}
\label{Data collection}
Tackling the GHG prediction problem from a disaggregated perspective, requires high resolution data points. To the best of our knowledge, currently there are no datasets available that report GHG at link level for a large and congested road network. An agent-based traffic and emission simulation has been used in this study to synthesize GHG and traffic characteristics for links at a very high temporal resolution  \cite{Djavadian2018Distributed}. Furthermore, we are of the view that in the near future, with the adoption of smart cities technologies, high resolution datasets from real road network would become available. In that case, the proposed LSTM framework can be retrained on such datasets. The agent based simulator used in this study implemented a calibrated Intelligent Driver Model (IDM) \cite{Treiber2000} for vehicular movement. Vehicles are dynamically routed on the network based on the real-time traffic information \cite{Djavadian2018Distributed}. Simulation ends once all of the vehicles reach their destinations. The link level space mean speed, density, and flow information is recorded every minute, which was found to be the optimal updating interval \cite{alfaseeh2018impact}. {Figure \ref{histograms}} demonstrates the statistical analysis of the variables reflecting on the traffic conditions and GHG ER (in $\text{CO}_{2\text{eq}}$ g/sec) on links. It can be noticed from {Figure \ref{histograms}} that the mode for speed is 40 km/h. It is crucial to note that the speed limit of 132 and 18 links in our case study is 60 and 80 km/h, respectively. Speed average in the network based on second-by-second data point is 56.16km/h. Nevertheless, speed varies from 0 to 80 km/h. Similarly, density (veh/km.lane) and flow (veh/h) are associated with a wide range reflecting on the different traffic conditions. Finally, GHG ERs (in $\text{CO}_{2\text{eq}}$) start from less than 1 (g/sec) to more than 5 (g/sec).

With regards to the emission modelling, Motor Vehicle Emission Simulator (MOVES), which is developed by the USEPA, is adopted to generate GHG ERs (in $\text{CO}_{2\text{eq}}$ g/sec) \cite{MOVES}. MOVES estimates emissions by defining the vehicle operating mode, which is based on the vehicle specific power (VSP) as illustrated in {Equation \ref{equ:1}}. 
\begin{equation}
	P_{V,t}=\frac{Av_t + Bv_t^2 + Cv_t^3+mv_ta_t}{m}
	\label{equ:1}
\end{equation}
\par Where: 
\begin{itemize}
	\item [] $P_{V,t}$ is the vehicle specific power (VSP) at time $t$
	\item [] $v_t$ is the speed of vehicle at time $t$ ($m/sec$)
	\item [] $a_t$ is the acceleration of vehicle at time $t$ ($m/sec^2$)
	\item [] $m$ is the mass of vehicle, usually referred as ``weight" (Mg). 
	\item[] A, B and C are track-road coefficients, representing rolling resistance, rotational resistance and aerodynamic drag, in unit kW-sec/m, kW-sec$^2$/$m^2$ and kW-sec$^3$/m$^3$.
\end{itemize}
The second-by-second CO$_2$ emissions of every vehicle on every link are then used to estimate the space mean GHG ER (in $\text{CO}_{2\text{eq}}$ g/sec) of each link based on a defined updating interval.

In terms of the data points, a variety of scenarios have been simulated to trigger different traffic conditions on links as shown in Table \textbf{\ref{scenarios investigated}}. 

\begin{table}[!h]
	\caption{Scenarios considered for data generation}
	\label{table1}
	\begin{center}
		\small
		\begin{tabular}[!t]{l c c}
			\hline
			Demand factor & No. of Vehicles & Departure time distribution \\
			\hline
			\hline
			\centering
			0.7 & 2,437  & Exponential, uniform, and normal\\
			\hline
			1 & 3,477 & Exponential, uniform, and normal  \\
			\hline
			1.3 & 4,520 & Exponential, uniform, and normal  \\
			\hline
			1.5 & 5,259 & Exponential \\
			\hline
			2 & 6,988 & Exponential \\
			\hline
		\end{tabular}
	\end{center}
	\label{scenarios investigated}
\end{table}

Travel demand is obtained from the Transportation Tomorrow Survey (TTS). The time dependent exogenous demand Origin-Destination (OD) matrices are based on 5 minute intervals from TTS. 
With regards to the demand, it ranges from 2,437 to 6,988 reflecting on demand factors from 0.7 to 2. To assure more heterogeneity in the traffic conditions generated at link level, different distributions are considered for the departure time, normal, uniform, and exponential. Two datasets are extracted corresponding to two time intervals i.e. 30 seconds and 1 minute, to examine the impact of the two levels of resolution on the prediction performance. The data has been pre-processed to suit each of the models discussed in Section \ref{Methodology} and to assure a fair comparison between the models. When 1 minute is the updating interval of predictors for clustering and LSTM, 48,652 and 12,159 data points for training and testing are employed, respectively. For the 30 seconds analysis the number is double. The data have been divided into 80\% training and 20\% testing for LSTM. With regards to clustering, 70\% of the data is randomly selected as the training set, 10\% of the total data is taken for validation, and 20\% of the data is used for testing, as demonstrated in {Figure \ref{methods}}. Finally, for the ARIMA model, 4 representative links are considered, which are associated with different characteristics to give an indication of the prediction performance at the network level.

\section{Results and discussion}
\label{Discussion and results}
This section presents the major findings, starting with the detailed correlation analysis outcome. Then we present the results of the ARIMA, clustering, and LSTM models as well as a comparison between them.

\subsection{Correlation analysis}
\label{Correlation_analysis_discussion}
\begin{figure*}[!h]
	\centering
	\includegraphics[width=5in]{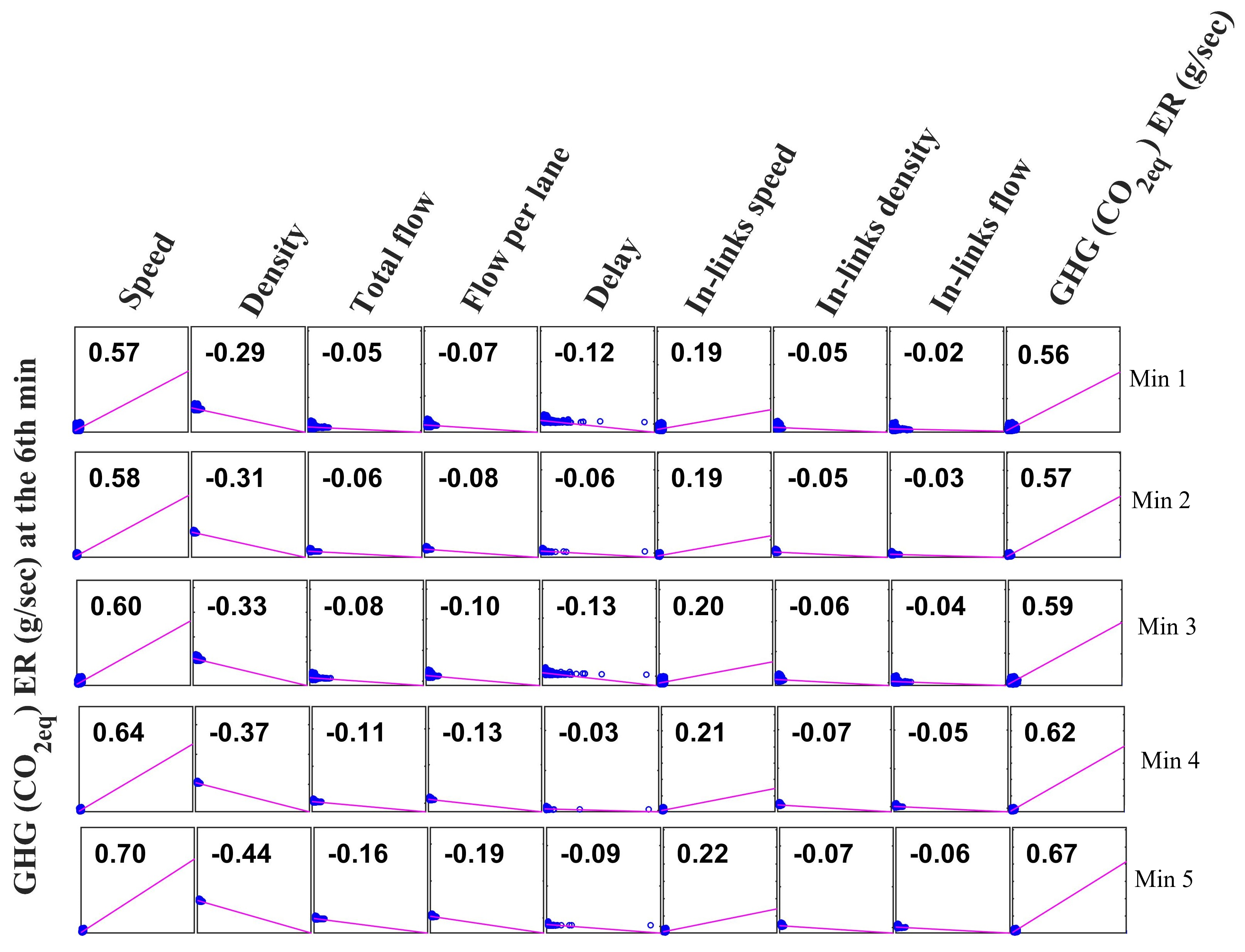}
	\caption{Correlation between the proposed variables at every minute from 1 to 5 (from top to bottom) with GHG ER (in $\text{CO}_{2\text{eq}}$ g/sec) at the 6th minute}
	\label{Correlation_1_to_5_with_6thminGHG_including_Speed}
\end{figure*}

The correlation analysis is the judging factor to define not only the most important predictors for the three models, but also the optimal number of sequences/minutes for the predictors while applying LSTM. Traffic and environmental information of studied links are captured. In order to expand the spatial dimension, the characteristics of in-links (upstream links) are not neglected. Traffic conditions (e.g. speed flow, density, etc.) at time $t$ on the upstream links will give a strong indication of the traffic condition on the studied link downstream at time $t+1$. The variables considered for this analysis are speed, density, flow, delay (difference between free flow travel time and actual travel time), in-links speed, in-links density, in-links flow, and GHG ER (in $\text{CO}_{2\text{eq}}$ g/sec). Five sequences/minutes have been investigated, which are sufficient to capture the changes in traffic conditions on link level in a congested urban network. The main outcome of this analysis is the correlation between the aforementioned indicators and the GHG ER (in $\text{CO}_{2\text{eq}}$ g/sec) at the sixth sequence/minute. Linear correlation has been employed for this part of the analysis. With reference to the results, the absolute value of the correlation coefficient between any variable and the GHG ER (in $\text{CO}_{2\text{eq}}$ g/sec) at the 6th min, except for delay, increases from minute 1 to minute 5 (top to down) as shown in {Figure \ref{Correlation_1_to_5_with_6thminGHG_including_Speed}}.

In terms of the importance order, speed, GHG ER (in $\text{CO}_{2\text{eq}}$ g/sec), density, and in-links speed are the top four highly correlated variables with the GHG ER at the 6th minute. Speed is the variable of the highest correlation coefficient with the GHG ER. This is due to the explicit dependency of GHG estimation on speed \cite{MOVES}. Based on the traffic fundamental relationships, the relationship between speed and density is observed to be monotonically decreasing, which justifies the high correlation between density and the GHG ER \cite{papacostas1993transportation}. Among the in-links characteristics, in-links speed is the most correlated variable with the GHG ER, which is due to the fact that speed is the variable GHG depends on for estimation. This analysis triggers the choice of not only the predictors for the three models, but also the number of sequences for LSTM.

\subsection{ARIMA with exogenous variables}
\label{ARIMAX_discussion}

\begin{figure*}[h!]
	\centering
	\includegraphics[width=2.8in]{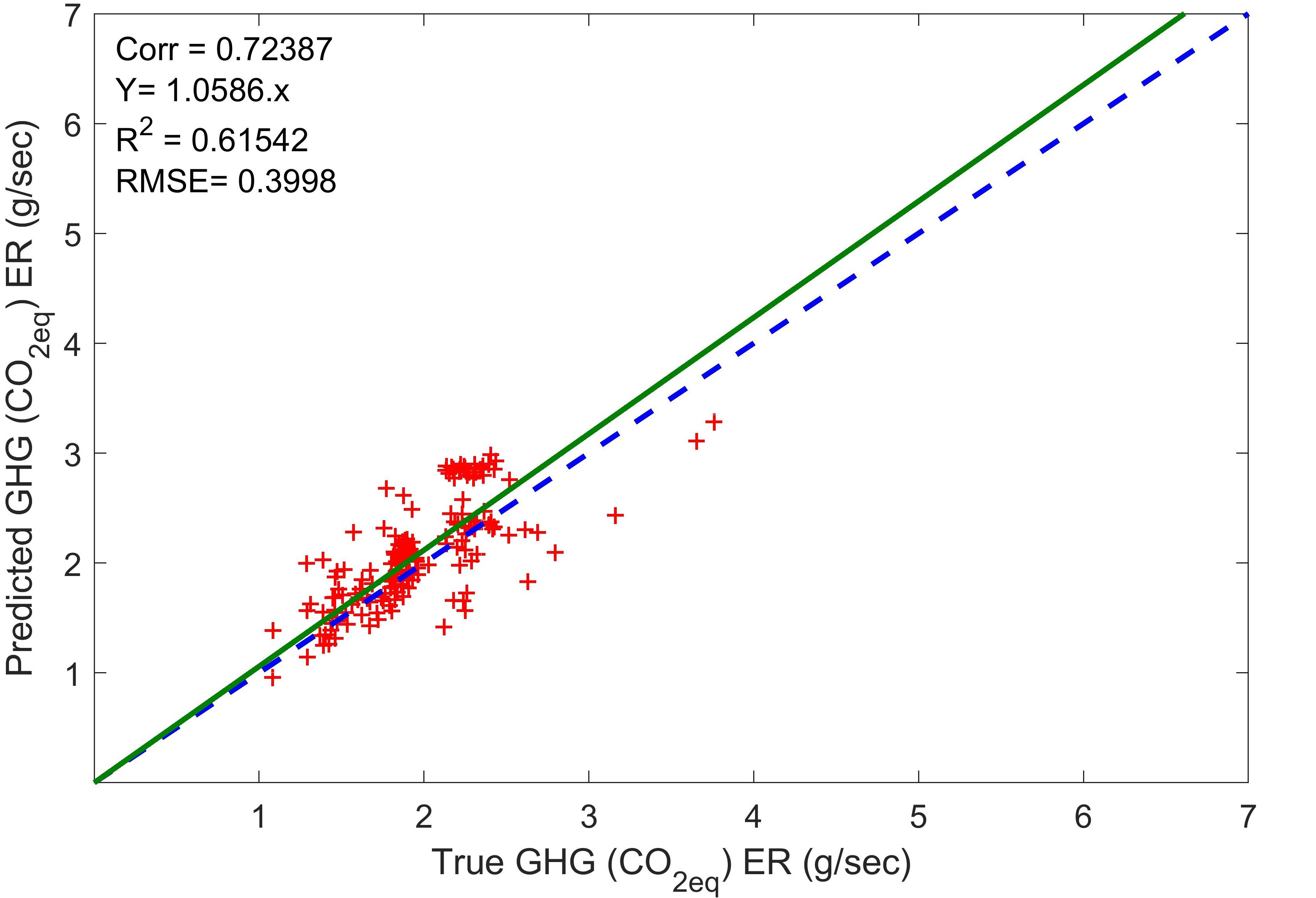}
	\caption{Sample representation (4 sampled links) of network level prediction using ARIMA with exogenous variables}
	\label{ARIMA_link_20_24_71_185_DF2_1min}
\end{figure*}

The ARIMA model with exogenous variables does not consider the non-linearity between variables \cite{zhang2003time}, which is a serious drawback when the data and relationships are complex. Furthermore, ARIMA model is not scaleable in which every link requires a model based on its data, which is tedious while dealing with a large number of links. Nevertheless, to compare between the models, a sample of 4 links, which are associated with different characteristics (number of lanes and free flow speed) and conditions (congested and uncongested), are considered. As shown in {Figure \ref{ARIMA_link_20_24_71_185_DF2_1min}}, the correlation coefficient and R$^2$ of the ARIMA model of the four links are 0.72 and 0.62, respectively. With regards to the fit to the straight curve, there is a slight overestimation and the RMSE is 0.3998 (g/sec).

\subsection{Clustering}
\label{Clustering_discussion}

\begin{figure*}[ht]
	\centering
	\includegraphics[width=3in]{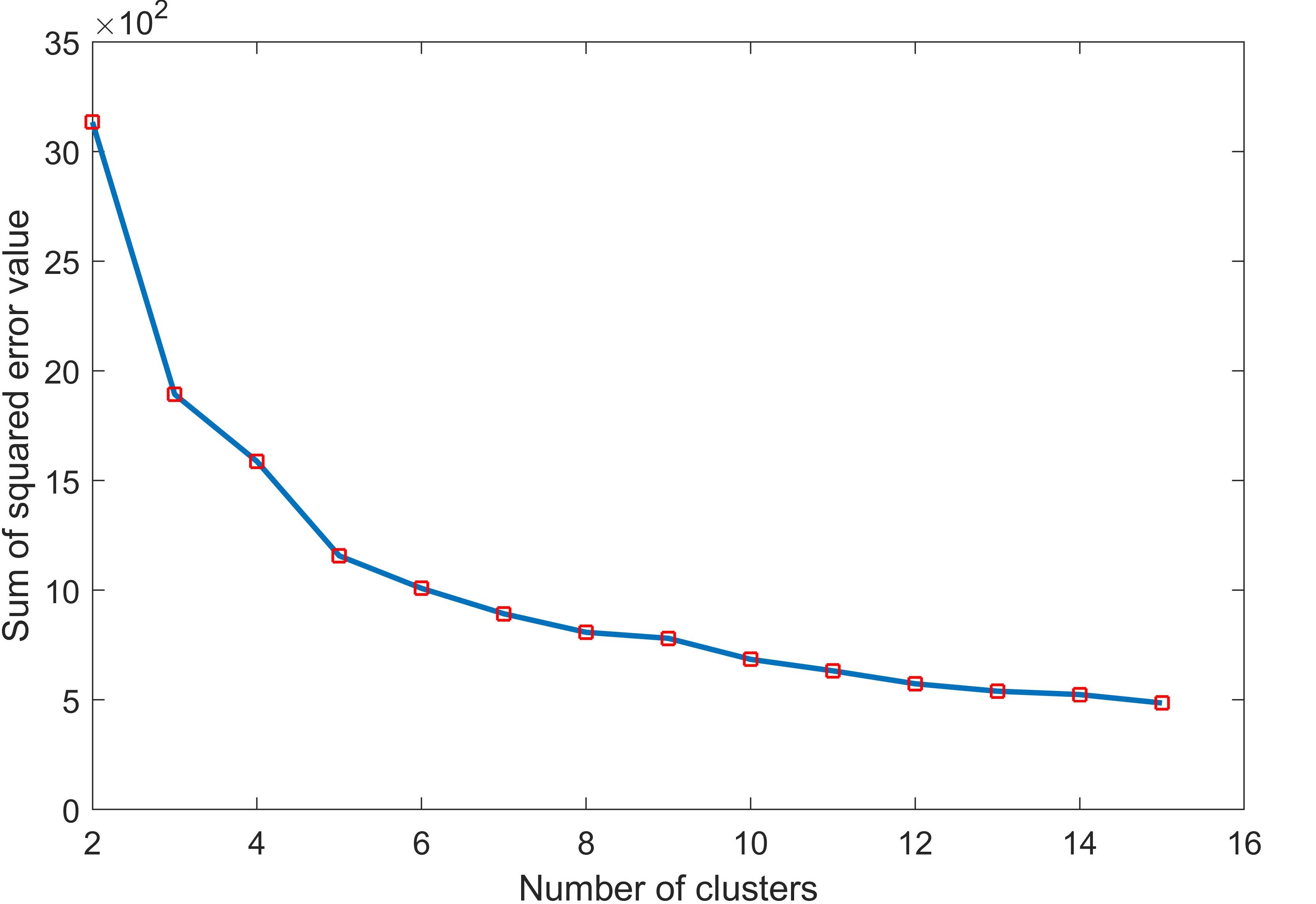}
	\caption{Sum of squared error (elbow method) of 15 clusters}
	\label{Sum_squared_error_final_15clusters_30sec_3pred}
\end{figure*}

\begin{figure}[!ht]
	\centering
	\begin{subfigure}[t]{.45\textwidth}
	{\includegraphics[width=3in]{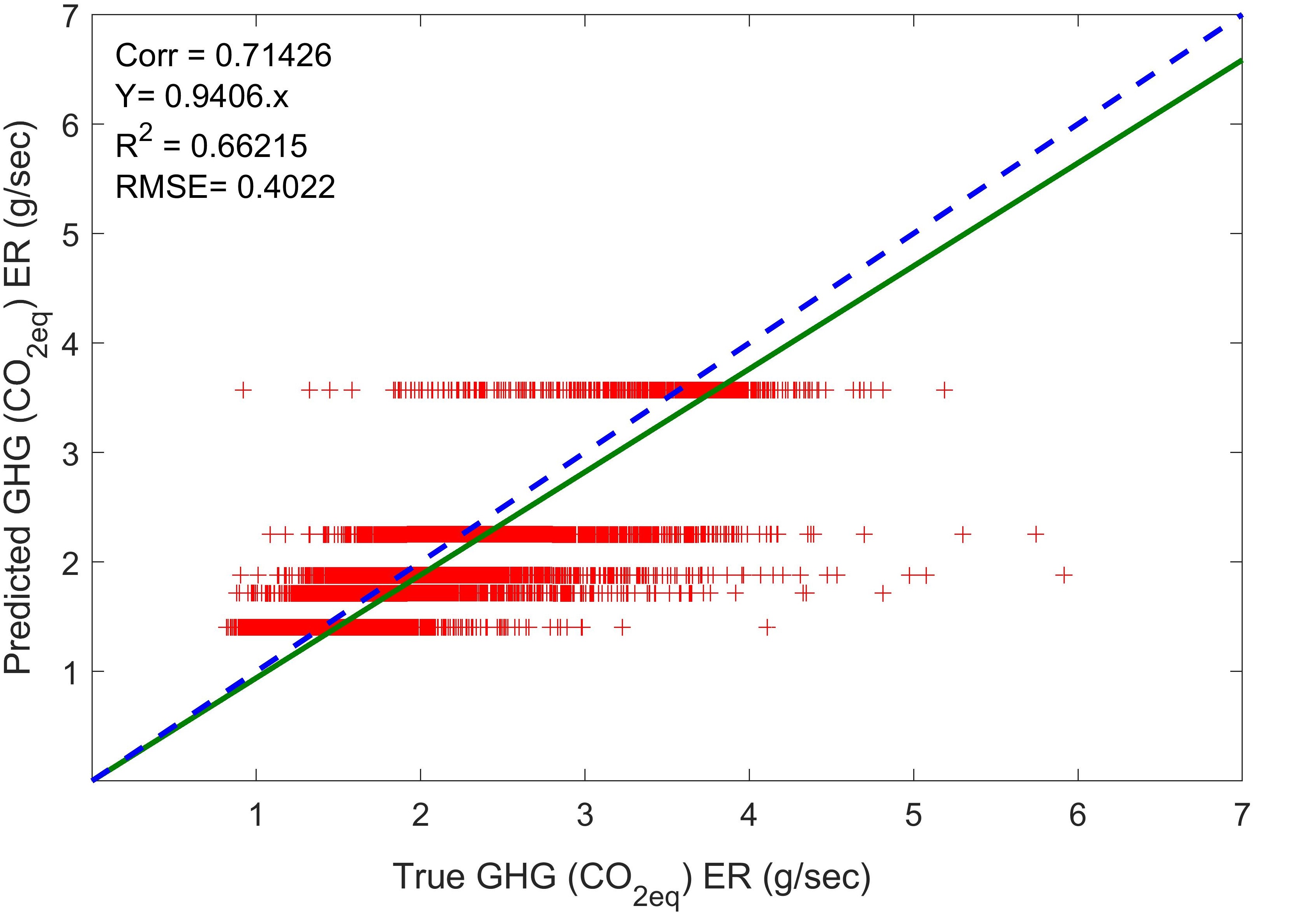}
		\label{5_clusters}}
	\end{subfigure}
	\hfil
	\begin{subfigure}[t]{.45\textwidth}
	{\includegraphics[width=3in]{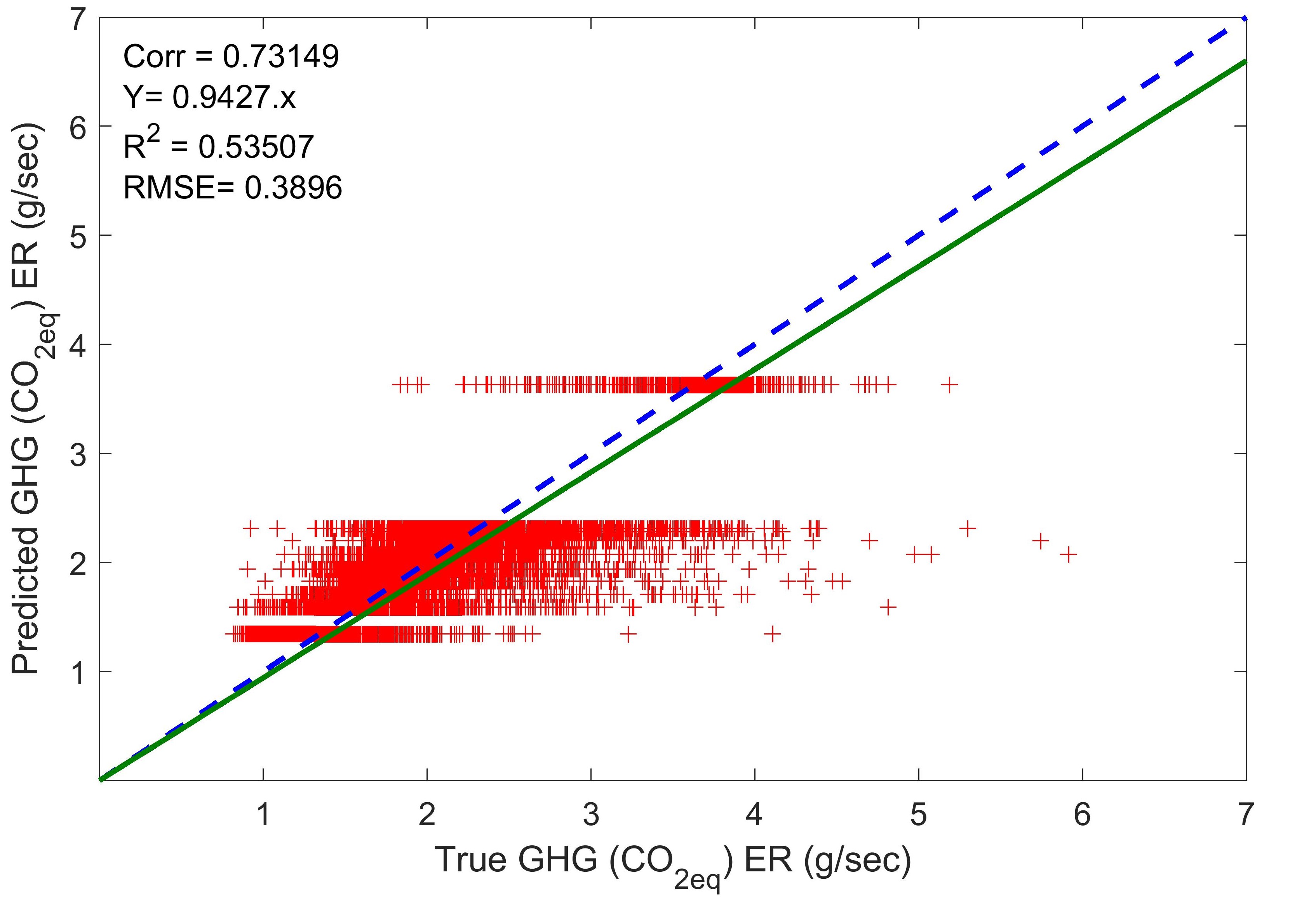}%
		\label{10_clusters}}
	\end{subfigure}
	\hfil
	\begin{subfigure}[t]{.45\textwidth}
		{\includegraphics[width=3in]{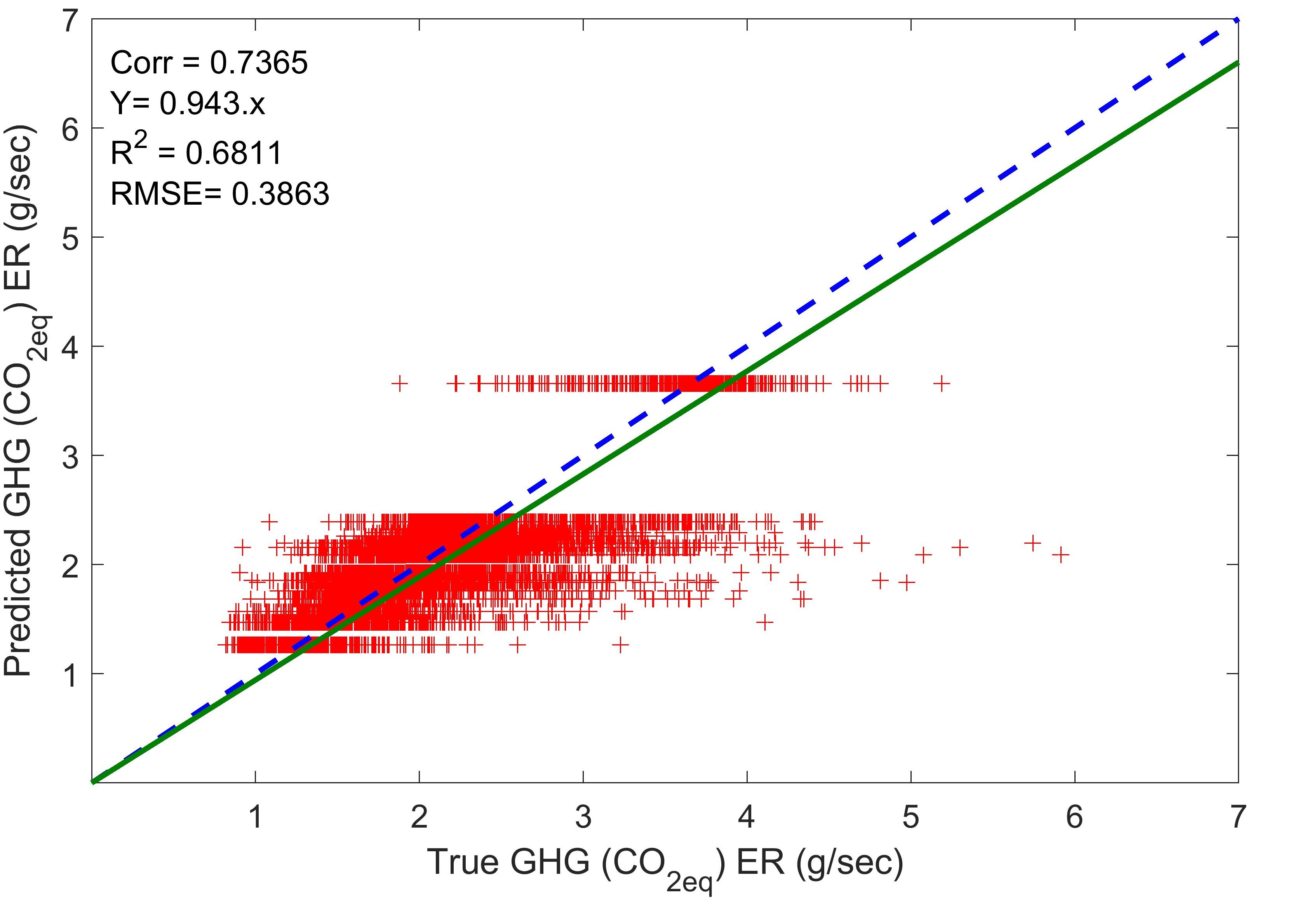}%
		\label{15_clusters}}
	\end{subfigure}
	\hfil
	\caption{True vs. predicted GHG ERs (in $\text{CO}_{2\text{eq}}$ g/sec) of clustering where: a) is for 5 clusters, b) is for 10 clusters, and c) is for 15 clusters}
	\label{Clustering}
\end{figure}

To define the optimal number of clusters, the sum of squared error (elbow method) is adopted. From {Figure \ref{Sum_squared_error_final_15clusters_30sec_3pred}}, the optimal number of clusters is 5. However, 10 and 15 clusters are examined for further enhancements. From Figure \ref{Clustering}, it is noticed that predicted values of GHG ERs are considered as discrete variables triggering dramatic over and underestimation. Values of GHG ERs ranging from 1 to 4 are predicted to be either 1.5, 1.8, 1.9, or 2.3 (g/sec). The correlation coefficient, R$^2$, and RMSE are 0.71, 0.66, and 0.4 (g/sec), respectively. While comparing true versus predicted GHG ERs of 5 clusters to 10 and to 15 clusters, it is illustrated that even when the number of clusters increases, the performance indicators have not improved substantially and the GHG ERs between 2.5 and 4 g/sec are either underestimated to be 2.5 or overestimated to be 4 g/sec.

With reference to the fit to the straight curve, it is noticed in Figure \ref{Clustering} that GHG ERs between 2.7 to around 4 (g/sec) are either underestimated taking one value of around 2.5(g/sec) or over estimated. It can be concluded that since clustering considers the predicted GHG ER as a discrete variable, the dynamic nature of traffic conditions is not captured efficiently, compared to ARIMA, triggering the aforementioned outcomes. Not to forget that there is a proportional relationship between the number of clusters and the amount of data required for training. Thus, even when the number of clusters is increased the data requirement can be hard to fulfill.

\subsection{LSTM with exogenous variables results}
\label{Long-Short Term Memory_discussion}
{Table \ref{LSTM scenarios investigated}} shows the specifications of the trained networks including the predictors, number of sequences (previous minutes), number of LSTM layers, updating interval of data points, and the hyper-parameter tuning approach.

\begin{table}[!ht]
	\begin{center}
		\small
		\begin{tabular}[!t]{c c c c c c c}
			\hline
			Model & Predictors & No. of &No. of& Sequence & Hyper-param.\\
			ID& & sequences/\ & LSTM & interval & tuning\\
			& & minutes & layers && approach\\
			\hline
			\hline
			\centering
			LSTM1 & Speed, density, & 3 &1 &1 min& Bayesian \\
			& and GHG ER (in $\text{CO}_{2\text{eq}}$ g/sec)&&\\
			\hline
			LSTM2 & Speed, density, GHG ER (in $\text{CO}_{2\text{eq}}$ g/sec), & 3 &1 &1 min& Bayesian\\
			& and in-links speed &&\\
			\hline
			LSTM3 & Speed, density, GHG ER (in $\text{CO}_{2\text{eq}}$ g/sec), & 3 & 2& 1 min& Bayesian\\
			& and in-links speed &&\\
			\hline
		\end{tabular}
	\end{center}
	\caption{LSTM specifications considered for application}
	\label{LSTM scenarios investigated}
\end{table}

Before demonstrating the results, it is important to note that the networks presented in this section are systematically tuned adopting the Bayesian optimization \cite{krizhevsky2009learning}. The search range for each of the hyper-parameters is around the optimal values of the manual tuning for each of the networks. For the manual tuning, a comprehensive list of tuning sets is examined to narrow the search range for a more efficient systematic tuning. Two solvers are considered, the stochastic gradient descent with momentum (sgdm) \cite{robert2014machine} and adaptive moment estimation (Adam) methods \cite{kingma2014adam}. The sgdm adds a momentum term to the parameter update to reduce oscillation associated with the standard stochastic gradient descent solver along the path of steepest descent towards the optimal solution \cite{robert2014machine}. Adam is derived from adaptive moment estimation. It adopts different learning rates for the parameters unlike sgdm, but uses a momentum term like the sgdm to further improve network training \cite{kingma2014adam}.

\begin{figure}[!h]
	\centering
			\begin{subfigure}[t]{.45\textwidth}
	{\includegraphics[width=3.2in]{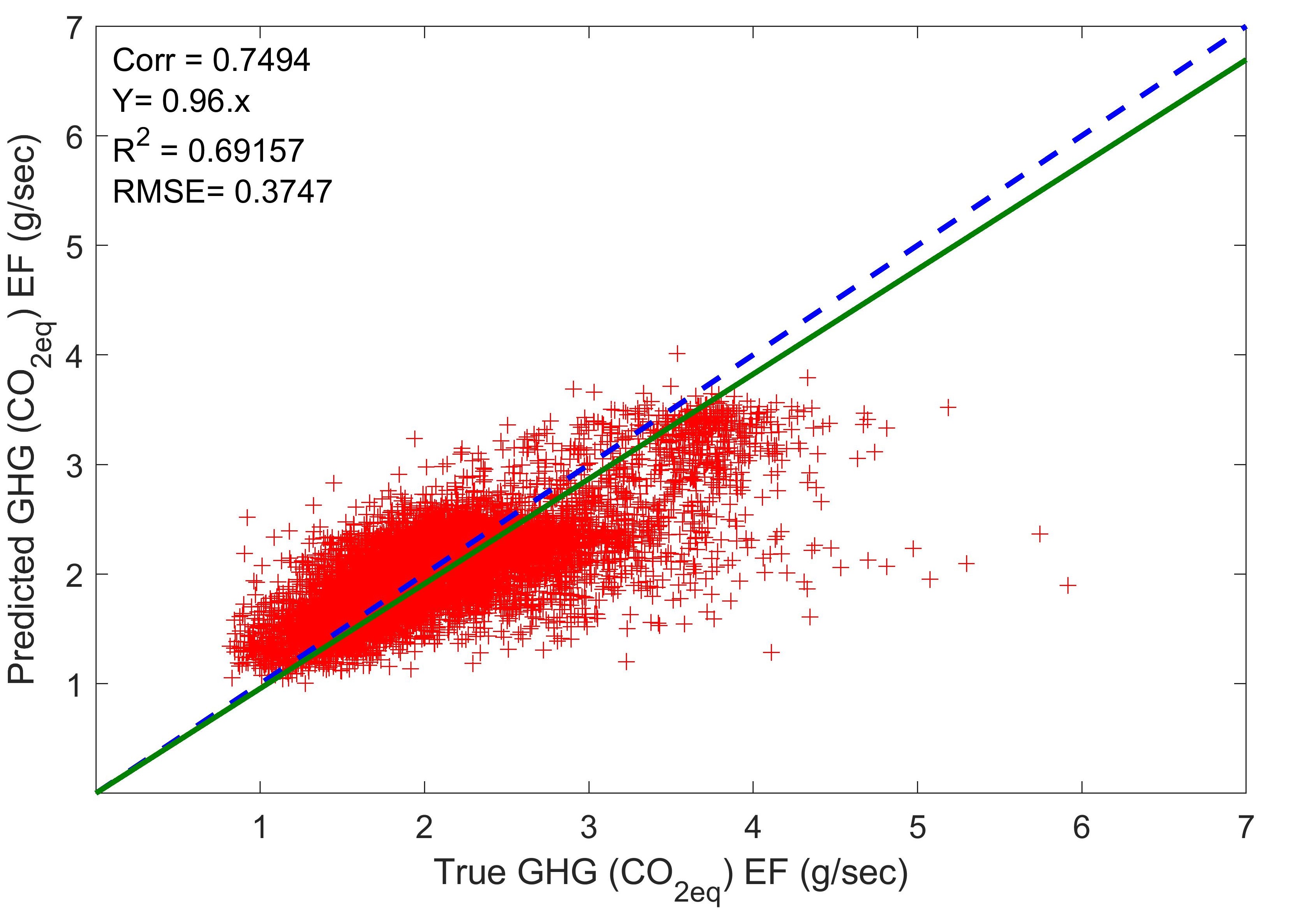}%
		\label{Iter10_7}}
	\end{subfigure}
	\hfil
			\begin{subfigure}[t]{.45\textwidth}
	{\includegraphics[width=3.2in]{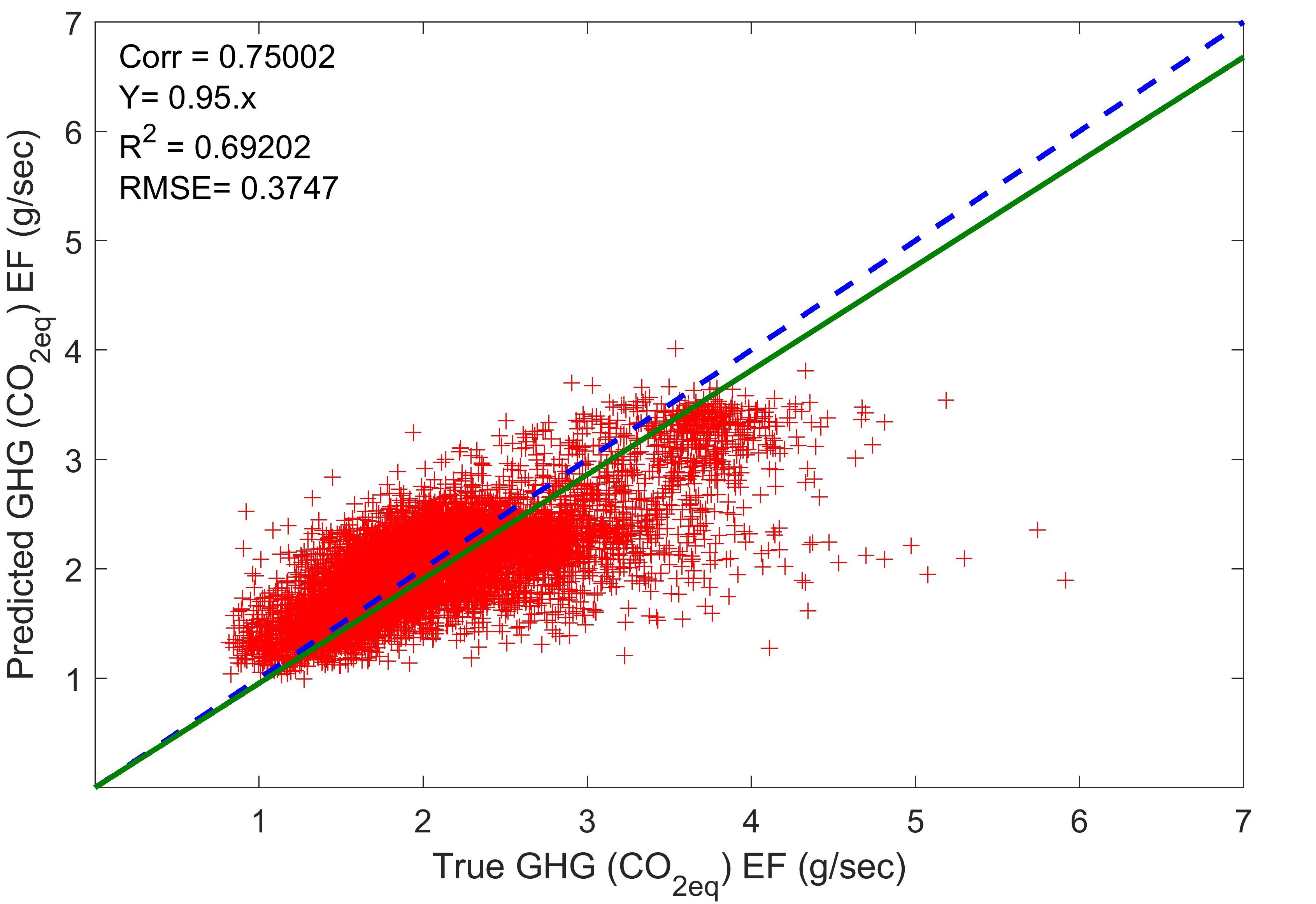}%
		\label{Iter11_7}}
	\end{subfigure}
	\hfil
			\begin{subfigure}[t]{.45\textwidth}
	{\includegraphics[width=3.2in]{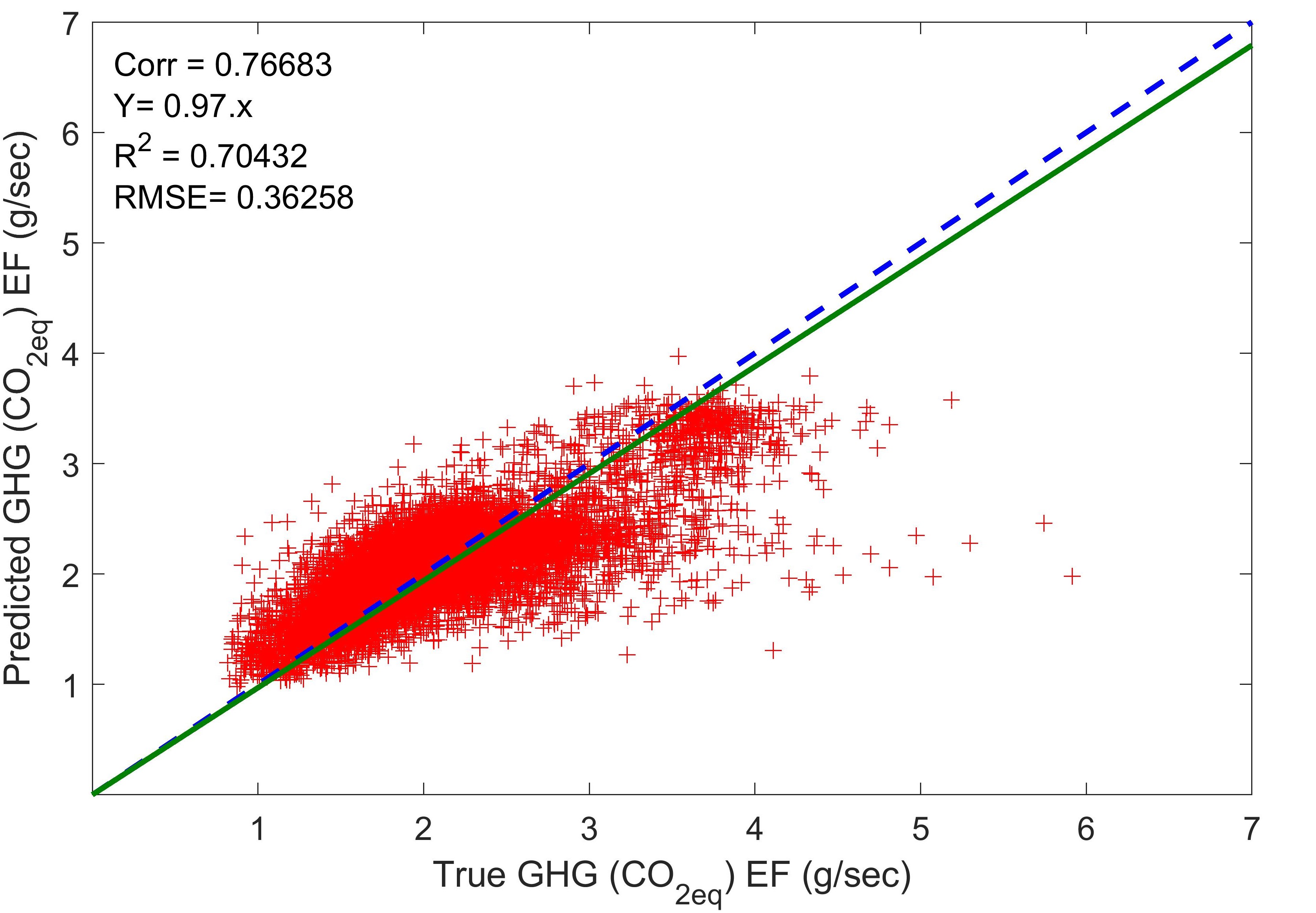}%
		\label{Iter11_10_3_2}}
	\end{subfigure}
	\hfil
	\caption{True vs. predicted GHG ERs (in $\text{CO}_{2\text{eq}}$ g/sec) of a) LSTM1, b) LSTM2, and c) LSTM3}
	\label{3LSTM_networks_final}
\end{figure}

The considered hyper-parameters include: the initial learning rate, momentum, max epochs, learning rate drop factor, learning rate drop period, number of hidden units of the first LSTM (hidden) layer, and the number of hidden units of the second (LSTM) layer when used. Not to neglect that different number of sequences (previous minutes) have been assessed for the best application. Three previous minute sequences have contributed to the best performance as the period is sufficient to capture the changes in traffic conditions and according to the linear correlation analysis as in Figure \ref{Correlation_1_to_5_with_6thminGHG_including_Speed}. {The performance indicators of the best LSTM networks are presented in Table \ref{Performance indicators of LSTM networks}.} Figure \ref{3LSTM_networks_final} demonstrates the performance indicators of LSTM1, which employs the link's three highly correlated predictors of the last three minutes. The LSTM1 is associated with a correlation coefficient, R$^2$, and RMSE of 0.75, 0.69, and 0.37 (g/sec), respectively, while underestimating the GHG ERs by 4\%. Comparing LSTM2 to LSTM1, a negligible enhancement in the correlation coefficient and R$^2$ is achieved while the RMSE is similar. The LSTM2 underestimates by 5\%. This slight improvement stems from the fact that LSTM2 does not only employ the three mostly correlated predictors with the response (GHG ER), but also the in-links highly correlated characteristic (in-links speed) as illustrated in Figure \ref{Correlation_1_to_5_with_6thminGHG_including_Speed}. When speed of in-links (upstream) at time $t$ is very low due to congestion contributing to more GHG emissions, this means that at time $t+1$ these vehicles driving at low speed will be on the downstream studied link. In other words, speed on in-links at time $t$ provides an indication of what the speed will be on the downstream link at time $t+1$. The results of LSTM2 demonstrates the potential of including the highly correlated in-links characteristics to better reflect on the spatial dimension while predicting GHG ERs. LSTM3, which adopts 2 hidden layers to account for the higher order of correlation between the variables, gives the best results in terms of the four performance indicators. The LSTM3 network underestimates by only 3\% and is associated with a correlation coefficient, R$^2$, and RMSE of 0.78, 0.7, and 0.36 (g/sec), respectively. 

\begin{table}[!ht]
	\begin{center}
		\small
		\begin{tabular}[!t]{c c c c c}
			\hline
			Model & Correlation & R$^2$ & Linear & RMSE \\
			ID & coefficient &  &  fit & \\
			\hline
			\hline
			\centering
			LSTM1 & 0.749 & 0.691 & Y=0.96.x & 0.374 \\
			\hline
			LSTM2 & 0.750 & 0.692 & Y=0.95.x & 0.374\\
			\hline
			LSTM3 & 0.767 & 0.704 & Y=0.97.x & 0.362\\
			\hline
		\end{tabular}
	\end{center}
	\caption{Performance indicators of the trained LSTM networks}
	\label{Performance indicators of LSTM networks}
\end{table}

A mutual feature in LSTMs is that the models are unable to predict GHG ERs that are higher than 4 g/sec. This is due to the fact that data points reflecting on traffic conditions associated are not sufficient as illustrated in Figure \ref{histograms}. That is, more data points representing the traffic conditions associated with GHG ERs higher than 4 (g/sec) may contribute positively. 

\par Investigating the impact of a shorter updating interval is of an added value. Hence, three LSTM models of the same specifications of LSTM1, LSTM2, and LSTM3 have been developed for 30 second intervals. The best model resulted in RSME of 0.48 g/sec, R$^2$ of 0.707, and correlation of 0.78. It is found that increasing the level of resolution has an adverse impact on the RMSE, while improving the correlation coefficient slightly for the three networks. More specifically, for LSTM3 when the updating interval is 30 second both the correlation coefficient and the RMSE increased, by 1.34\% and 35\%, respectively. The main justification is that the shorter updating interval triggers more noise and local fluctuations that are not well captured by a general LSTM that is not trained for a specific link. Therefore, 1 minute updating interval is sufficient at this point to account for the high level of heterogeneity in link characteristics.

\par Comparing the LSTM model outcome to the other models, clustering and ARIMA, shows that the three aforementioned LSTM networks outperform the best case of clustering as in Figure \ref{Clustering} and ARIMA as in Figure \ref{ARIMA_link_20_24_71_185_DF2_1min} in terms of the four performance indicators. More specifically, the RMSE of LSTM3 is less by 6\% and 9\% compared to the best case of clustering and ARIMA, respectively. Unlike clustering, LSTM considers GHG ER (in $\text{CO}_{2\text{eq}}$ g/sec) as a continuous variable. Compared to ARIMA with exogenous variables, LSTM is scaleable and is applicable to any link in the studied network. It is important to note that the links in downtown Toronto are associated with a high level of heterogeneity in terms of the speed limit and the number of lanes. Thus, the developed LSTM models are network level regardless of the link characteristics, which might contribute to the deterioration in prediction performance when GHG ERs are higher than 4 g/sec.

\section{Conclusion and potential directions}
\label{Conclusion}
It has been shown that the efficiency of routing is a key tool to reduce the undesirable impact of transportation systems on the environment \cite{tu2019quantifying}. Thus, reliably predicting GHG ER (in $\text{CO}_{2\text{eq}}$ g/sec) is crucial. It can pave the way to develop non-myopic eco-routing {resulting in more sustainable transportation systems} and can also be adopted in other applications that can efficiently mitigate the negative effects of GHG emissions. Previous studies are predominantly based on macroscopic data points. Even when microscopic data are used, the case studies involve a single or few intersections/links \cite{alfaseeh2020multifactor}. Therefore, there is a need for studies tackling the issues related to GHG emissions at a high resolution in terms of both time and space. To predict GHG ERs at link level, a deep learning approach, i,e, LSTM with exogenous variables has been employed in this study. LSTM has been chosen because it has overcome various limitations of other NNs in the context of time-series data, such as the vanishing gradient problem \cite{amarpuri2019prediction}. To assure satisfying prediction outcome, a sufficient amount of representative data points is generated from MOVE and traffic microsimulation for downtown Toronto. The optimal set of predictors and number of time sequences used are identified based on a comprehensive correlation analysis. A comparison with the commonly used approaches for times series prediction, i.e. clustering and ARIMA, is also conducted.

In this study, the LSTM3 network of two hidden layers, which is systematically tuned using Bayesian optimization, outperforms the best case of clustering and ARIMA models in terms of the four performance indicators. More specifically, the LSTM3 network reduces the RMSE by 6\% and 9\% compared to the best case of clustering and ARIMA, respectively. While the LSTM can scale up to a network level and considers the GHG ER as a continuous variable, ARIMA requires a model for every link and clustering considers GHG ER as a discrete variable. ARIMA's major drawbacks are related to scaling and non-linearity consideration between variables. In clustering, the main limitation is that predicted GHG ER (in $\text{CO}_{2\text{eq}}$ g/sec) is considered as a discrete variable, which has affected the fit quality to the straight curve. In other words, GHG ERs (in $\text{CO}_{2\text{eq}}$ g/sec) between 2.5 and 4 g/sec are either underestimated to be 2.5 or overestimated to be 4, even when the number of clusters is 15. This is due to the fact that clustering lacks the temporal dimension compared to the LSTM model. Utilizing a smaller updating interval of 30 seconds in the case of LSTM modelling improves the correlation coefficient slightly compared to the case of 1 minute updating interval, but has a negative impact on the RMSE of the three LSTM models. The shorter updating interval triggers more local fluctuations that are not well captured by the networks level model.

For future work, various applications of the model will be evaluated. One such application is in the non-myopic eco-routing, where future time-step GHG predictions on downstream links are needed in order to decide the dynamic route for a vehicle. Figure \ref{histograms} illustrates that the frequency of speed higher than 60 (km/h) or lower than 30 (km/h) is very small. Hence, to enhance the current work, generating more data points reflecting on the aforementioned condition may contribute positively to the current outcomes. With reference to the hyper-parameters tuning, dedicating more time for the Bayeisan optimization may result in further improvement of the model's prediction capacity. Developing models for categorized links based on their speed limit, and number of lanes may introduce further enhancements. In addition, utilizing physical constraints, which consider suitable traffic characteristics (flow over capacity, density over jam density, etc.), may contribute to higher prediction accuracy. Finally, since the traffic and environmental information are obtained from simulation, employing sensors to collect real data will further enhance the model's prediction capability.





\bibliographystyle{abbrv}
\bibliography{references.bib}
\end{document}